\documentclass[aps,prb,preprint, citeautoscript,floatfix,superscriptaddress]{revtex4-2}

\usepackage{braket}
\usepackage{bm}
\usepackage{graphicx}
\usepackage{amsmath}
\usepackage{amssymb}
\usepackage{colordvi}
\usepackage{multirow}

\usepackage[hidelinks]{hyperref}
\usepackage{color}
\usepackage{ulem}

\begin{document}

\title{Gate voltage induced injection and shift currents in AA- and AB-stacked bilayer graphene}

\author{Ze Zheng}
\affiliation{GPL Photonics Laboratory, State Key Laboratory of Luminescence and Applications, Changchun Institute of Optics, Fine Mechanics and Physics, Chinese Academy of Sciences, Changchun, Jilin, 130033 P. R. China}
\affiliation{University of Chinese Academy of Sciences, Beijing 100039, China.} 
 
\author{Kainan Chang}
\email{knchang@ciomp.ac.cn}
\affiliation{GPL Photonics Laboratory, State Key Laboratory of Luminescence and Applications, Changchun Institute of Optics, Fine Mechanics and Physics, Chinese Academy of Sciences, Changchun, Jilin, 130033 P. R. China}
\affiliation{University of Chinese Academy of Sciences, Beijing 100039, China.} 

\author{Jin Luo Cheng}
\email{jlcheng@ciomp.ac.cn}
\affiliation{GPL Photonics Laboratory, State Key Laboratory of Luminescence and Applications, Changchun Institute of Optics, Fine Mechanics and Physics, Chinese Academy of Sciences, Changchun, Jilin, 130033 P. R. China}
\affiliation{University of Chinese Academy of Sciences, Beijing 100039, China.} 

\date{\today}

\begin{abstract}
Generating photogalvanic effects in centrosymmetric materials can provide new opportunities for developing passive photodetectors and energy harvesting devices. In this work, we investigate the photogalvanic effects in centrosymmetric two-dimensional materials, AA- and AB-stacked bilayer graphene, by applying an external gate 
voltage to break the symmetry. Using a tight-binding model to describe the electronic states, the injection coefficients for circular photogalvanic effects and shift conductivities for linear photogalvanic effects are calculated for both materials with light wavelengths ranging from THz to visible. 
We find that gate voltage induced photogalvanic effects can be very significant for AB-stacked bilayer graphene, 
with generating a maximal dc current in the order of mA for a 1 $\mu$m wide sample illuminated by a light intensity of 0.1 GW/cm$^2$, which is determined by the optical transition around the band gap and van Hove singularity points.
Although such effects in AA-stacked bilayer graphene are about two orders of magnitude smaller than those in AB-stacked bilayer graphene, the spectrum is interestingly limited in  a very narrow photon energy window, which is associated with the interlayer coupling  strength. 
A detailed analysis of the light polarization dependence is also performed. The gate voltage and chemical potential can be used to effectively control the photogalvanic effects.
\end{abstract}
\maketitle

\newpage
\section{Introduction}
  Photogalvanic effects are nonlinear optical responses that
  generate direct currents in homogeneous materials, and such a passive
  process is considered as a direct and powerful photoelectric
conversion method~\cite{NatureMaterials_18_471_2019_Osterhoudt,npjComputationalMathematics_9_35_2023_Sauer,NaturePhotonics_10_611_2016_Spanier}. 
The widely discussed photogalvanic effects can be induced by the one-color injection current and shift current, which are second order nonlinear optical processes occurring in noncentrosymmetric materials, or the two-color coherent current injection processes, which are third (for ``1+2'' process)~\cite{Phys.Rev.B_93_75442_2016_Salazar} or fifth  (for ``2+3'' process)~\cite{Phys.Rev.Lett._123_67402_2019_Wang} order nonlinear optical processes and are not sensitive to the inversion symmetry of materials.
According to the response to the light polarization, second order photogalvanic effects are also
  phenomenologically divided into circularly
polarized
photogalvanic effect and linearly polarized photogalvanic effect, where the
  latter is light phase insensitive and can be used for solar energy
  harvest without forming p-n junctions to surpass the Shockley-Queisser
  limit~\cite{JournalofAppliedPhysics_32_510_1961_Shockley,Kaner2020,NatureCommunications_8_14176_2017_Cook}. 
 One of
the research topics in this field is to find materials with significant
photogalvanic effects at a specific frequency range, and several studies have been conducted on various new materials, including 2D
materials~\cite{Phys.Rev.B_104_115402_2021_Wei,XunCM2021,
YuanNn2014,Phys.Rev.B_104_241404_2021_Arora,
PhysicalReviewResearch_4_13209_2022_Kaplan}, Dirac or Weyl semimetals~\cite{NatureMaterials_18_471_2019_Osterhoudt,NatureMaterials_18_955_2019_Ji,Phys.Rev.B_95_41104_2017_Chan}, ferroelectric materials~\cite{Phys.Rev.Lett._109_116601_2012_Young,ScientificReports_8_8005_2018_Pal,AngewandteChemie_132_3961_2020_Peng,AdvancedMaterials_22_1763_2010_Ji}, and so on.

As the first two-dimension material, graphene is a potential
  candidate for realizing new functionality in optoelectronic devices
  due to its superior optical and electronic properties exceeding many
  traditional bulk materials. However, because of its centrosymmetric
  crystal structure, one-color injection and shift currents vanish in
  many few-layer graphene as well as their nanostructures, while
  two-color coherent control has been well studied in both theories~\cite{Phys.Rev.B_61_7669_2000_Mele,Phys.Rev.B_91_235320_2015_Cheng,Phys.Rev.B_93_75442_2016_Salazar,ZhengPRB2022}
  and experiments~\cite{sun2010coherent,sun2012current}. It is still meaningful to generate one-color injection and shift currents in centrosymmetric graphene based
  structure, in order to utilize its extraordinary physical
  properties. The generation of second order response can be realized
  by forming an asymmetric interface or edge~\cite{Phys.Rev.B_99_165432_2019_Vandelli}, applying an external
  electric field~\cite{Phys.Rev.B_91_205405_2015_Brun}, forming surface curvature~\cite{lin2014observation}, considering the spatial
  variation of the light field~\cite{cheng2017second}, and stacking graphene layers into asymmetric structure~\cite{Scienceadvances_4_74_2018_Shan}. 
Wei \textit{et al.}~\cite{Phys.Rev.B_104_115402_2021_Wei} studied the gate field induced injection and
  shift currents in zigzag graphene nanoribbons, and found
  that the subband and edge states determine the generated currents
  with an effective modulation of their amplitudes by the ribbon width and the
static field strength.
 Xiong \textit{et al.}~\cite{2DMaterials_8_35008_2021_Xiong} investigated the light polarization dependence
  of in-plane shift current  in a AB-stacked bilayer
  graphene~(AB-BG) with applying a gate voltage, and their results
  clearly illustrated a sizeable photocurrent at a given light
  frequency; however, neither the spectra
  of the shift conductivity nor the injection current was present.
By stacking two layers of monolayer graphene with a relative rotation to form a twisted bilayer graphene, a large shift current can be produced due to a huge density of states when the flat band is formed at magic angles~\cite{PhysicalReviewResearch_4_13209_2022_Kaplan,Phys.Rev.B_104_241404_2021_Arora,Phys.Rev.Res._4_13164_2022_Chaudhary}.  Surprisingly, whether the gate voltage can generate
  photogalvanic effect in AA-stacked bilayer graphene~(AA-BG) is still not clear.

In this paper, we systematically study the spectra of the
  injection coefficients and shift conductivities of AA-BG and AB-BG under applying a gate voltage to break the
  inversion symmetry, as well as their dependence on the gate voltage
  and chemical potential. 
Their electronic states are described
by widely adopted tight-binding model
formed by the carbon $2p_z$ orbitals
\cite{Phys.Rev.B_91_205405_2015_Brun,JournalofthePhysicalSocietyofJapan_81_124713_2012_Chuang},
and the expressions for injection coefficient and shift conductivity
are employed from Ref.\,[\citenum{Phys.Rev.B_61_5337_2000_Sipe}].
Our results confirm the feasibilities of generating photogalvanic effects in AA-BG and AB-BG. Particularly, the response of AA-BG distributes in a very narrow spectral region, 
while  a maximal  current in the order of mA  can be generated in AB-BG for a 1 $\mu$m wide sample at light intensity of 0.1 GW/cm$^2$.

This paper is organized as follows. 
In Sec.\,\uppercase\expandafter{\romannumeral2} we introduce the
tight-binding models for the AA-BG and AB-BG under applying a gate voltage, and give  the expressions for the
injection coefficient and shift conductivity. 
In Sec.\,\uppercase\expandafter{\romannumeral3} we present the spectra of injection coefficient and shift
  conductivity for  AA-BG and AB-BG,
and discuss the effects of the gate voltage and chemical potential. 
We conclude in Sec.\,\uppercase\expandafter{\romannumeral4}.

\section{Models}
\subsection{Hamiltonian}

\begin{figure*}[ht]
	\centering
	\includegraphics[scale=1,trim= 40 0 50 0,clip]{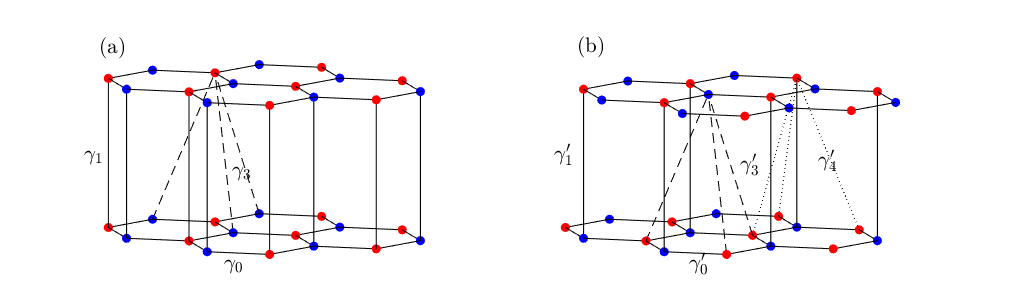}
	\caption{Crystal structures and tight-binding hopping parameters for (a) AA-BG and (b) AB-BG.}
	\label{fig:structure}
\end{figure*}

We consider the tight-binding Hamiltonian  for the AA-BG and AB-BG, whose crystal structures are
illustrated in Fig.\,\ref{fig:structure} (a) and (b), respectively.
These two structures have the same primitive lattice vectors $\bm
a_1=a_0\left(\frac{1}{2}\hat{\bm x} + \frac{\sqrt{3}}{2} \hat{\bm y}\right)$
and $\bm a_2=a_0\left(-\frac{1}{2}\hat{\bm x} + \frac{\sqrt{3}}{2} \hat{\bm y}\right)$ with the lattice constant
$a_0=2.46$ {\AA}. 
The atomic positions in the unit cell are taken as
$\bm\tau_A=\bm 0$, $\bm\tau_B=(\bm a_1+\bm a_2)/3$,
$\bm\tau_{A'}=c\hat{\bm z}$, and $\bm\tau_{B'}=\bm\tau_B+c\hat{\bm z}$ for
AA-BG, and  $\bm \tau_A=\bm 0$, $\bm\tau_B=(\bm a_1+\bm a_2)/3$, $\bm
\tau_{A'}=\bm\tau_B+c\hat{\bm z}$, and $\bm \tau_{B'}=2\bm
\tau_B+c\hat{\bm z}$ for AB-BG, where 
$c=3.35$ {\AA}  is the interlayer distance. 
The primitive reciprocal lattice vectors are  $\bm
b_1=\frac{2\pi}{a_0}\left(\hat{\bm x} + \frac{1}{\sqrt{3}}\hat{\bm
    y}\right)$ and $\bm b_2=\frac{2\pi}{a_0}\left(-\hat{\bm
    x}+\frac{1}{\sqrt{3}}\hat{\bm y}\right)$. 
The electronic states are described by a tight-binding model employing carbon $2p_z$ orbitals.  
The unperturbed Hamiltonian \cite{JournalofthePhysicalSocietyofJapan_81_124713_2012_Chuang} for
AA-BG is
\begin{align}
	H^{\rm AA}_{\bm k}=\left(
	\begin{array}{cccc}
		-\Delta & \gamma_0g_{\bm k}& \gamma_1 & \gamma_3g_{\bm
                                                        k} \\
		\gamma_0g^*_{\bm k} & -\Delta & \gamma_3g^*_{\bm k} & \gamma_1\\
          \gamma_1 & \gamma_3g_{\bm k} & \Delta & \gamma_0g_{\bm k}\\
          \gamma_3g^*_{\bm k} & \gamma_1 & \gamma_0g^*_{\bm k} & \Delta\\
	\end{array}
	\right)\,.
	\label{H_AA}
\end{align}
Here $\bm k$ is the electron wavevector, and $g_{\bm
    k}=1+e^{-i\bm k\cdot \bm a_1}+e^{-i\bm k\cdot \bm a_2}$. The
hopping parameters are illustrated in Fig.\,\ref{fig:structure} (a) with
$\gamma_0=2.569$~eV,  $\gamma_1=0.361$~eV, and
$\gamma_3=-0.032$~eV. 
The on-site energies $\pm\Delta$ are induced by
a gate voltage.
The Hamiltonian for AB-BG is given from Ref.\,\citenum{Phys.Rev.B_91_205405_2015_Brun} as
\begin{align}
	 H^{\rm AB}_{\bm k}=\left(
	\begin{array}{cccc}
		-\Delta-\frac{\Delta'}{2} & \gamma_0^\prime g_{\bm k}&
                                                                      \gamma_4^\prime
                                                                      g_{\bm
                                                                      k}
          & \gamma_3^\prime g^*_{\bm k} \\
		\gamma_0^\prime g^*_{\bm k} &
                                              -\Delta+\frac{\Delta'}{2} & \gamma_1^\prime & \gamma_4^\prime g_{\bm k}\\
		\gamma_4^\prime g^*_{\bm k} & \gamma_1^\prime &
                                                                \Delta+\frac{\Delta'}{2} & \gamma_0^\prime g_{\bm k}\\
		\gamma_3^\prime g_{\bm k} & \gamma_4^\prime g^*_{\bm
                                            k} & \gamma_0^\prime g^*_{\bm k} & \Delta-\frac{\Delta'}{2}\\
	\end{array}
	\right)\,,
	\label{eq:H_AB}
\end{align}
where the hopping parameters (see Fig.\,\ref{fig:structure} (b)) are $\gamma_0^\prime=-3.16$~eV,
  $\gamma_1^\prime=0.381$~eV, $\gamma_3^\prime=-0.38$~eV, and $\gamma_4^\prime=0.14$~eV. 
  The on-site
  potential difference $\Delta^\prime=0.022$~eV is induced by the
  asymmetric environment of A, B atoms in the crystal structure.

The eigenstates $C_{n\bm k}$ and eigenenergies $\epsilon_{n\bm k}$ at the $n$th band
are obtained by diagonalizing the Hamiltonian through
\begin{align}
  H_{\bm k}C_{n\bm k}=\epsilon_{n\bm k}C_{n\bm k}\,.
\end{align}
The calculation of the optical responses involves the position operator $\widetilde{\bm r}_{\bm k}$ and velocity
  operator $\widetilde{\bm v}_{\bm k}$, which are 
\begin{align}
		\widetilde{\bm r}_{\bm k} &= i\bm\nabla_{\bm k} +
		\left(\begin{array}{cccc}
		 \bm\tau_A & 0 & 0 & 0 \\ 0 & \bm\tau_B &
		0 & 0 \\ 0 & 0 & \bm\tau_{A'} &0 \\ 0 & 0&0&
                \bm\tau_{B'}
                \end{array}\right)\,,\quad 
	 \widetilde{\bm v}_{\bm k}=\frac{1}{i\hbar}[ \widetilde{\bm r}_{\bm k}, H_{\bm k}]\,,
\end{align}  
respectively.
The matrix elements of the position operator give the Berry
connections $\bm\xi_{nm\bm k}$ by
\begin{align}
  \bm\xi_{nm\bm k}=C^{\dagger}_{n\bm k}\widetilde{\bm r}_{\bm k}C_{m\bm k}\,,	\label{eq:xi}
\end{align}
and those of the velocity operator are calculated as $\bm v_{nm\bm k}=C^{\dagger}_{n\bm k}\widetilde{\bm v}_{\bm k}C_{m\bm k}$.
Due to the derivative with respect to the wavevector $\bm k$, a direct calculation of $\bm\xi_{nm\bm k}$ from Eq.\,(\ref{eq:xi}) requires that the wavefunction $C_{n\bm k}$ is a
  smooth function of $\bm k$.
However, this becomes quite difficult in numerical calculation because of
the phase arbitrary for a numerical wavefunction. Practically,  the off-diagonal
terms of $\bm\xi_{nm\bm k}$ can be also calculated from the velocity
operator as
\begin{align}
	\bm r_{nm\bm k}=\begin{cases}
		\bm\xi_{nm\bm k}=\frac{\bm v_{nm\bm k}}{i\omega_{nm\bm k}} &  (n\ne m)\\
		0         &  (n=m)
	\end{cases} \,,
\end{align}
with $\hbar\omega_{nm\bm k}=\epsilon_{n\bm k}-\epsilon_{m\bm k}$. 
The diagonal terms $\xi^a_{nn\bm k}$ usually appear in the  generalized derivative of $(r^c_{\bm k})_{;nmk^a}=\frac{\partial r_{nm\bm k}^c}{\partial k^a}-i(\xi^a_{nn\bm k}-\xi^a_{mm\bm k})r^c_{nm\bm k}$, which is calculated alternatively~\cite{Phys.Rev.B_104_115402_2021_Wei} by
\begin{align}
  (r^c_{\bm k})_{;nmk^a}=&\frac{-ir^c_{nm\bm k}{\cal V}_{mn\bm k}^a+\hbar M^{ca}_{nm\bm
                           k}+i[r^a_{\bm k}, v^c_{\bm
                           k}]_{nm}}{i\omega_{nm\bm k}}\,,
\end{align}
with ${\cal V}_{mn\bm k}^a=v^a_{mm\bm
                           k}-v^a_{nn\bm k}=\frac{\partial
                           \omega_{mn\bm k}}{\partial k^a}$ and 
\begin{align}
	M^{ca}_{nm\bm k}=C^{\dagger}_{n\bm k}\frac{1}{i\hbar}[\widetilde{r}^a_{\bm k},\widetilde{v}^c_{\bm k}]C_{m\bm k}\,,
\end{align}
where the Raman letters $a,c$ indicate the Cartesian directions
$x,y,z$. Note that the electron wavevector has only in-plane
  components $x,y$, the derivative $\frac{\partial}{\partial k^z}$
  thus gives zero and $\left(r_{\bm
    k}^a\right)_{;nmk^z}=-i(\xi^z_{nn\bm k}-\xi^z_{mm\bm k})r^a_{nm\bm
  k}$.

\subsection{Injection and shift currents}
We focus on the injection and shift currents induced by
  a laser pulse centered at frequency
    $\omega$, for which the electric field is $\bm E(t)=\bm
  E_0(t)e^{-i\omega t}+c.c.$ and $\bm E_0(t)$ is
a slow varying envelop function. 
The response static currents can be written as 
\begin{align}
\bm J_0(t)=\bm J_{\rm inj}(t)+\bm J_{\rm sh}(t)\,.
\label{eq:total-current}
\end{align}
Here the first term $\bm J_{\rm inj}(t)$ is a one-color injection current satisfying
\begin{align}
	\frac{dJ^a_{\rm inj}(t)}{dt}=2i\eta^{abc}(\omega)E_0^b(t)[E_0^c(t)]^*\,,
\end{align}
with the injection coefficient $\eta^{abc}(\omega)$ given by
\begin{align}
  \eta^{abc}(\omega)
  =\frac{2e^3\pi}{\hbar^2}\int\frac{d\bm
  k}{4\pi^2}\sum_{nm}{\cal V}_{mn\bm k}^af_{nm\bm
  k} \text{Im}[r^c_{mn\bm k}r^b_{nm\bm k}]\delta(\omega_{mn\bm k}-\omega)\,.
    \label{eq:inj}
\end{align}
Here $f_{nm\bm k}=f_{n\bm k}-f_{m\bm k}$ is the population difference
with the Fermi-Dirac distribution $f_{n\bm k}=[1-e^{(\epsilon_{n\bm
    k}-\mu)/k_BT}]^{-1}$ for given chemical potential $\mu$ and
temperature $T$.
The second term $\bm J_{\rm sh}(t)$ in Eq.\,(\ref{eq:total-current})  is a shift current written as
\begin{align}
	J^a_{\rm sh}(t)&=2\sigma^{abc}(\omega)E_0^b(t)[E_0^c(t)]^*\,,
\end{align}
with the shift conductivity $\sigma^{abc}(\omega)$ given by
\begin{align}
  \sigma^{abc}(\omega)
  &=-\frac{i\pi e^3}{\hbar^2}\int\frac{d\bm
    k}{4\pi^2}\sum_{nm}f_{nm\bm k}\left[r^b_{mn\bm k}\left(r_{\bm k}^c\right)_{;nmk^a}+r^c_{mn\bm
    k}\left(r_{\bm k}^b\right)_{;nmk^a}\right]\delta(\omega_{mn\bm  k}-\omega)\,.
    \label{eq:sh}
\end{align}

Further discussion of photocurrents starts with a
symmetry analysis on the tensors of $\eta^{abc}(\omega)$ and $\sigma^{abc}(\omega)$. 
The presence of time-reversal symmetry 
gives $\bm r_{nm\bm k}=\bm r_{mn(-\bm k)}=[\bm r_{nm (-\bm k)}]^*$,
$\bm v_{nm\bm k}=-\bm v_{mn(-\bm k)}=-[\bm v_{nm(-\bm k)}]^*$, $\epsilon_{n \bm k}=\epsilon_{n (-\bm k)}$, and $\left(r^b_{\bm k}\right)_{;nm
  k^a}=-\left(r^b_{-\bm k}\right)_{;mn k^a}=-[\left(r^b_{\bm k}\right)_{;nmk^a}]^*$. Thus from
Eqs.\,(\ref{eq:inj}) and (\ref{eq:sh}), we obtain $\eta^{abc}=[\eta^{abc}]^*$ and
$\sigma^{abc}=[\sigma^{abc}]^*$, which are both real numbers.
At finite gate voltage, the crystal point group of AB-BG is
  $C_{3v}$, whose symmetry is lower than that of AA-BG with crystal
  point group $C_{6v}$. Thus we can check the symmetry properties of
   AB-BG first, and then refine them to AA-BG. Combining the point
  group and the time reversal symmetry, the nonzero tensor components
  satisfy $\eta^{xzx}=\eta^{yzy}=\eta^{xxz}=\eta^{yyz}$,   $\sigma^{xzx}=\sigma^{yzy}=\sigma^{xxz}=\sigma^{yyz}$,
  $\sigma^{zxx}=\sigma^{zyy}$, $\sigma^{zzz}$, and
  $\sigma^{yyy}=-\sigma^{yxx}=-\sigma^{xxy}=-\sigma^{xyx}$.  Then the
  injection current becomes
\begin{align}
  \frac{dJ_{\rm inj}^a(t)}{dt}=4\eta^{xzx}(\omega)
 \text{Im}\{E^a_0(t)[E^z_0(t)]^\ast\}(1-\delta_{a,z})\,,
	\label{eq:inj_curren}
\end{align}
and the shift current is
\begin{subequations}
  \label{eq:shift_curren}
\begin{align}
  J_{\rm sh}^x(t)&=4\sigma^{xzx}(\omega)\text{Re}\left\{E^z_0(t)[E^x_0(t)]^*\right\}-4\sigma^{yyy}(\omega)\text{Re}\left\{E^x_0(t)[E^y_0(t)]^*\right\}\,,\\
	J_{\rm
  sh}^y(t)&=4\sigma^{xzx}(\omega)\text{Re}\left\{E^z_0(t)[E^y_0(t)]^*\right\}+2\sigma^{yyy}(\omega)
         \left[ |E^y_0(t)|^2-|E^x_0(t)|^2\right]\,,	\\
	J_{\rm sh}^z(t)&=2\sigma^{zxx}(\omega)\left[|E^x_0(t)|^2+|E^y_0(t)|^2\right]
	+2\sigma^{zzz}(\omega)|E^z_0(t)|^2\,.
\end{align}
\end{subequations}
For AA-BG, the results are similar except that the $\sigma^{yyy}$
component disappears due to the extra crystal symmetry. 

The injection current in AA-BG or AB-BG requires an elliptically
  polarized light incident obliquely, and its $z$-component vanishes
  due to the lack of freely moving electrons along this quantum
  confined direction. The $z$-component of shift current  in AA-BG or
  AB-BG,  induced by the charge shift
  between the two layers under the light excitation,  can be always  generated. Such shift current can lead to charge
  accumulation between these two layers, which can further induce a 
  gate voltage in this system, as discussed by Gao {\it et al.}
  \cite{Phys.Rev.Lett._124_77401_2020_Gao}. The in-plane components of
  the shift current in AA-BG can be generated only for  an
  elliptically polarized light incident obliquely, while  those in
  AB-BG have no such limit.

\section{Results}
\subsection{Analytical results for AA-BG}
The Hamiltonian for the AA-BG can be analytically diagonalized. The eigenstates are
\begin{align}
	C_{n\bm k}&=\frac{\sqrt{1-
			\alpha_n{\cal N}_{\beta_n\bm k}}}{2\sqrt{2}} \left(
	\begin{array}{c}
          -\hat{g}_{\bm k}\\
	- \beta_n\\
	\beta_n\hat{g}_{\bm k}\\
	1
\end{array}
\right) +\frac{\alpha_n\sqrt{1+\alpha_n{\cal N}_{\beta_n\bm k}}}{2\sqrt{2}}                       
	\left(
	\begin{array}{c}
          \hat{g}_{\bm k}\\
		\beta_n\\
		\beta_n\hat{g}_{\bm k}\\
		1
	\end{array}
	\right)\,,\label{eq:wfaa}
\end{align}
with $\hat{g}_{\bm k}={g_{\bm k}}/{|g_{\bm k}|}$ and 
\begin{align}
{\cal N}_{\beta_n\bm k}=\frac{\gamma_3|g_{\bm k}|+\beta_n\gamma_1}{\sqrt{\Delta^2+(\gamma_3|g_{\bm k}|+\beta_n\gamma_1)^2}}\,.
\end{align}
Here $n=1,2,3,4$ denotes the band index with
  $\alpha_n=-1,-1,+1,+1$ and $\beta_n=-1,+1,-1,+1$, respectively. The associated eigenenergies are
\begin{align}
  \epsilon_{n\bm k}
  =\beta_n\gamma_0|g_{\bm k}|+\alpha_n\sqrt{\Delta^2+(\gamma_3|g_{\bm k}|+\beta_n\gamma_1)^2}\,.
\end{align}

With the analytic wavefunctions in Eq.\,(\ref{eq:wfaa}), Berry
  connections $\bm \xi_{nm\bm k}$ can be calculated directly from Eq.\,(\ref{eq:xi}), as listed in Appendix~\ref{app:xi}, where the relations between all
components are also presented. There exist selection rules for
$r_{nm\bm k}^z$ as
\begin{align}
  r^z_{13\bm k}&=r^z_{31\bm k}=\frac{c{\cal
                 N}_{-1\bm k}}{2}\,, \quad 
                 r^z_{24\bm k}=r^z_{42\bm k}=\frac{c{\cal
			N}_{+1\bm k}}{2}\,.\label{eq:rz}
\end{align}
Therefore, $r_{nm\bm k}^z$ is nonzero only for the band pair
$(n,m)=(1,3)$  or $(2,4)$.  The injection coefficient becomes
\begin{align}
	\eta^{xzx}(\omega)=&\frac{e^3}{2\pi \hbar^2}\int d\bm
                             k\left\{f_{13\bm k}{\cal
                             V}_{31\bm k}^x\text{Im}[r^x_{31\bm k}r^z_{13\bm k}]\delta(\omega_{31\bm k}-\omega)
                             \right.\notag\\
  &\left.\hspace{2.2cm}+f_{24\bm k}{\cal V}_{42\bm k}^x\text{Im}[r^x_{42\bm k}r^z_{24\bm k}]\delta(\omega_{42\bm k}-\omega)\right\}\,.
	\label{eq:AA_inj}
\end{align}
The intraband Berry connections are obtained as
\begin{align}
 \bm \xi_{nn\bm k}
  &=\frac{1}{2}\left[g_{\bm k}^\ast(i\bm\nabla_{\bm k})g_{\bm
    k}+\frac{a_0}{\sqrt{3}}\hat{\bm y}\right]+\frac{1}{2}c\hat{\bm z}\left(1+\alpha_n\sqrt{1-{\cal
      N}_{{\beta}_n\bm k}^2}\right)\,,\label{eq:xinn}
\end{align}
The matrix elements for $\xi_{nn\bm
  k}^{x/y}$ are independent of the band index $n$, thus $(r_{\bm
  k}^a)_{;nmk^b}=\frac{\partial r_{nm\bm k}^a}{\partial k^b}$ for
$b=x,y$ and $(r_{\bm
  k}^a)_{;nmk^z}=-i(\xi_{nn\bm k}^z-\xi_{mm\bm k}^z)r_{nm\bm
  k}^a$. The shift conductivities become
\begin{subequations}
	\label{eq:AA_sigma}
\begin{align}
  \sigma^{xzx}(\omega)
  =&-i\frac{e^3}{4\pi\hbar^2}\int d\bm k \left[f_{13\bm k}\left(r^z_{31\bm k}\frac{\partial r^x_{13\bm k}}{\partial k_x}+r^x_{31\bm k}\frac{\partial r^z_{13\bm k}}{\partial k_x}\right)\delta(\omega_{31\bm k}-\omega)\right.\notag\\
   &\left.\hspace{2.8cm}+f_{24\bm k}\left(r^z_{42\bm k}\frac{\partial r^x_{24\bm k}}{\partial k_x}+r^x_{42\bm k}\frac{\partial r^z_{24\bm k}}{\partial k_x}\right)\delta(\omega_{42\bm k}-\omega)\right]\,,\label{eq:AA_sigmaxzx}\\
  \sigma^{zzz}(\omega)
  =&\frac{e^3}{2\pi\hbar^2}\int d\bm
     k\left[f_{12\bm k}|r^z_{31\bm
     k}|^2(\xi_{33\bm k}^z-\xi_{11\bm k}^z)\delta(\omega_{31\bm k}-\omega)\right.\notag\\
   &\left.\hspace{2.1cm}	+f_{24\bm k}|r^z_{42\bm k}|^2(\xi_{44\bm
     k}^z-\xi_{22\bm k}^z)\delta(\omega_{42\bm k}-\omega)\right]\,,\label{eq:AA_sigmazzz}\\
  \sigma^{zxx}(\omega)
  =&\frac{ e^3}{2\pi \hbar^2}\int d\bm
     k \sum_{nm} f_{nm\bm k}|r^x_{mn\bm
     k}|^2(\xi_{mm\bm k}^z-\xi_{nn\bm k}^z)\delta(\omega_{mn\bm  k}-\omega)\,,
\end{align}
\end{subequations}
It can be seen that  the coefficients $\eta^{xzx}$, $\sigma^{xzx}$, and
$\sigma^{zzz}$ are induced by the transitions only from the band 1 to 3 or
from the band 2 to 4, while $\sigma^{zxx}$ has no such
limit. 
  These coefficients can be further simplified with the analytical expressions of all these quantities, which can be obtained under the linear dispersion
  approximation around the Dirac points, as shown in
  Appendix~\ref{app:xiaa}.
  
\begin{figure*}[ht]
	\centering
	\includegraphics[scale=1]{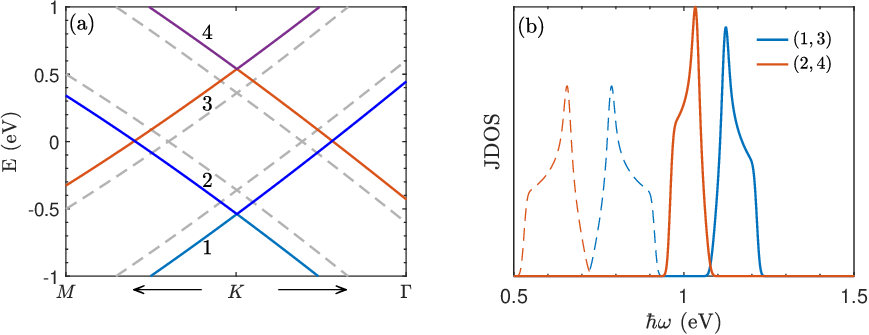}
	\caption{Band structure (a) and JDOS (b) for AA-BG at
          $\Delta=0$ (dashed curves) and $\Delta=0.4$~eV (solid curves).}
	\label{fig:band_AA}
\end{figure*}

Figure \ref{fig:band_AA}\,(a) shows the band structure of
AA-BG for $\Delta=0$ and $0.4$~eV. With applying a gate
  voltage, the interlayer coupling shifts the energies of the Dirac
  cones of each layer, while the electronic states at zero energy are
  still degenerate. The
bands 1 and 3 (or 2 and 4) are approximately parallel to each
other, and their energy differences 
are in the range of
$2\sqrt{\Delta^2+(\gamma_1+3\gamma_3)^2}\le\hbar\omega_{42\bm
  k}\le2\sqrt{\Delta^2+\gamma_1^2}\le\hbar\omega_{31\bm k}\le2\sqrt{\Delta^2+(\gamma_1-3\gamma_3)^2}$
due to $0\le|g_{\bm k}|\le 3$, where the middle value is obtained at the Dirac points and the
other two values are obtained at the M points. Figure
  \ref{fig:band_AA}\,(b) gives the joint density of states (JDOS)
  ${\cal J}_{31}(\omega)$ and ${\cal J}_{42}(\omega)$ for
  related two pairs of bands, which are defined as
  \begin{align}
    {\cal J}_{nm}(\omega) &= \int d\bm k \delta(\hbar\omega_{nm\bm k}-\hbar\omega)\,.
  \end{align}
  These two JDOS are strongly localized in energy, regardless of whether there is the gate voltage. For $\Delta=0.4$~eV, ${\cal D}_{42}(\omega)$ is nonzero in the energy range of $[0.95,
  1.08]$~eV and ${\cal D}_{31}(\omega)$  is nonzero in the energy range of $[1.08,
  1.21]$~eV. 

\subsection{Band structure of AB-BG }

\begin{figure*}[ht]
	\centering
	\includegraphics[scale=1]{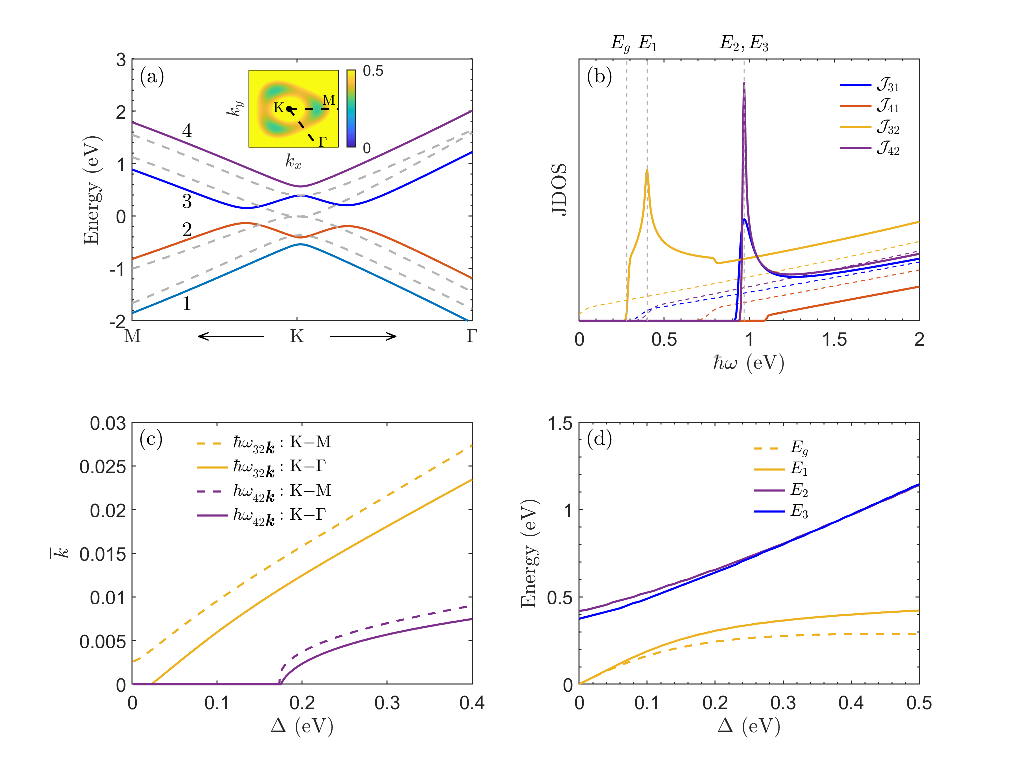}
          \caption{(a) Band structure and (b) JDOS  for AB-BG at
          $\Delta=0$ (dashed curves) and $\Delta=0.4$~eV (solid curves).             
The energetic locations of the band gap ($E_g$) and the maximal values of ${\cal J}_{32}$, ${\cal J}_{42}$, and ${\cal J}_{31}$ ($E_1$, $E_2$, $E_3$)  are indicated.
          The inset: $\bm k$-resolved energy difference $\hbar\omega_{32\bm
            k}$ for $\Delta=0.4$~eV.            
          (c)  $\Delta$ dependence of the  $\bm k$ location for the minimum of $\hbar\omega_{nm\bm k}$ along the K-M and K-$\Gamma$ directions.
          (d)  $\Delta$ dependence of  $E_g$,  $E_1$,  $E_2$, and $E_3$.}
	\label{fig:band_AB}
\end{figure*}

The Hamiltonian in Eq.\,(\ref{eq:H_AB}) for AB-BG can be also 
  analytically diagonalized, as shown in Appendix~\ref{app:ab}, but the
  expressions for the eigenenergies are too complicated to provide
  meaningful physical insight, thus we discuss the band structure based on numerical calculation.
 This work focuses on the electronic transitions around the Dirac points, for convenience, the wavevectors are expressed as 
 $\bm k=\bar{k}\frac{2\pi}{a_0} (\hat{\bm x}\cos\theta+\hat{\bm y}\sin\theta)+\bm K$
 with $\theta=2n\pi/3$ along the K-M directions, and $\theta=(2n+1)\pi/3$ along the K-$\Gamma$ directions. 
Figure \ref{fig:band_AB}\,(a) gives the
band structure for AB-BG at gate voltages $\Delta= 0$ and 0.4 eV.
At $\Delta=0$, in each Dirac cone, the two middle bands are degenerate at the Dirac points with $\bar{k}=0$  and other three $\bm k$ points on the K-M paths with $\bar{k}=
-\frac{\gamma_1^\prime\gamma_3^\prime}{\sqrt{3}\pi{\gamma_0^\prime}^2}
    \sim 0.003$ (see details in Appendix~\ref{app:ab}). 
  Meanwhile, the energy differences, $\hbar\omega_{31\bm k}$ and $\hbar\omega_{42\bm k}$, have minima at the Dirac points.
 For nonzero gate voltage, the degeneracy at these points is lifted.
 The eigenenergies at the Dirac points are  $\pm \Delta-\frac{\Delta^\prime}{2}$,
$\pm\sqrt{\Delta^2+\gamma_1^2}+\frac{\Delta^\prime}{2}$,
and the middle two bands around the Dirac points have the Mexican hat shape \cite{McCann2013}.
At $\Delta=0.4$ eV, the energy difference $\hbar\omega_{32\bm k}$ shows a minimum with increasing $\bar{k}$ for each $\theta$, as shown in the $\bm k$-resolved energy difference in the inset, where 
the three-fold rotational symmetry can be clearly seen around this Dirac point. 
Along the K-M directions, the minima  of $\hbar\omega_{32\bm k}$ appear around $\bar{k}=0.027$ to give the band gap of $E_g=0.28$ eV; and along the K-$\Gamma$ directions, the minima appear around $\bar k=0.023$, which have an energy $E_1=0.4$~eV higher than the band gap and give a van Hove singularity (VHS).  
Similar results can be found for $\hbar\omega_{42\bm k}$, and another VHS appears with energy $E_2=0.97$ eV; however, $\hbar\omega_{31\bm k}$ shows a minimum at the Dirac points but no VHS appears. 
Figure \ref{fig:band_AB}\,(b) gives JDOS of ${\cal J}_{31}(\omega)$, ${\cal J}_{32}(\omega)$, ${\cal J}_{41}(\omega)$, and ${\cal J}_{42}(\omega)$ at $\Delta=0$ and $0.4$ eV. 
The gate voltage changes these JDOS significantly around the band edge.
 ${\cal J}_{32}(\omega)$ and ${\cal J}_{42}(\omega)$ have divergences at the VHS points with energies $E_1$ and $E_2$, respectively; and ${\cal J}_{31}(\omega)$ has a peak located at $E_3\sim0.97$~eV around the band edge, 
which is induced by the nearly parallel bands (1, 3) around the Dirac points.

The VHS points do not appear for all gate voltages.
Figure \ref{fig:band_AB}\,(c) exhibits 
$\Delta$ dependence of the $\bar{k}$ value for the minimal energy of $\hbar\omega_{32\bm k}$ and $\hbar\omega_{42\bm k}$ for $\theta$ along the K-M and K-$\Gamma$ directions, respectively. 
Along the K-M directions, $\hbar\omega_{32\bm k}$ has a minimum value at nonzero $\bar{k}$ for all $\Delta$, which gives the band gap $E_g$ of the system; while along the K-$\Gamma$ directions, the minimum energy $E_1$ moves to a nonzero $\bar{k}$ only for $\Delta \geq 0.023$~eV, where VHS appears as well. Note that the JDOS ${\cal J}_{32\bm k}$ shows a maximum at the band edge when there is no VHS for $\Delta < 0.023$ eV. 
However, the minima of  $\hbar\omega_{42\bm k}$ along the K-M and K-$\Gamma$ directions
locate not at the Dirac points only for 
 $\Delta\geq0.174$ eV, where VHS appears as well. For $\Delta<0.174$ eV, ${\cal J}_{42}(\omega)$ also shows a maximum at the band edge between the bands 4 and 2, where this energy  is still noted as $E_2$; the maximum of ${\cal J}_{31}(\omega)$ also locates at the band edge between bands 3 and 1, where this energy is still noted as $E_3$. 
The gate voltage dependences of these energies $E_g$, $E_1$, $E_2$, and $E_3$ are shown in Fig.\,\ref{fig:band_AB}\,(d).

\subsection{Injection coefficients and shift conductivities at $\Delta=0.4$~eV}

In this section we present the numerical results for 
  injection coefficient $\eta^{xzx}(\omega)$ and shift
  conductivities $\sigma^{yyy}(\omega)$, $\sigma^{xzx}(\omega)$,
  $\sigma^{zxx}(\omega)$, and $\sigma^{zzz}(\omega)$. The parameters
  are chosen as $T=300$~K, $\mu=$ 0, $\Delta=0.4$~eV. During the numerical calculation, the Brillouin
zone is divided  into a $3000\times3000$
homogeneous grid.
The $\delta$ functions in Eqs.\,(\ref{eq:inj}) and (\ref{eq:sh}) are approximated by a Gaussian function as
$\delta(\omega)=\frac{\hbar}{\sqrt{\pi}\Gamma}e^{-(\hbar\omega)^2/\Gamma^2}$
with the Gaussian broadening $\Gamma=10$ meV.

\begin{figure*}[ht]
	\centering
	\includegraphics[scale=1]{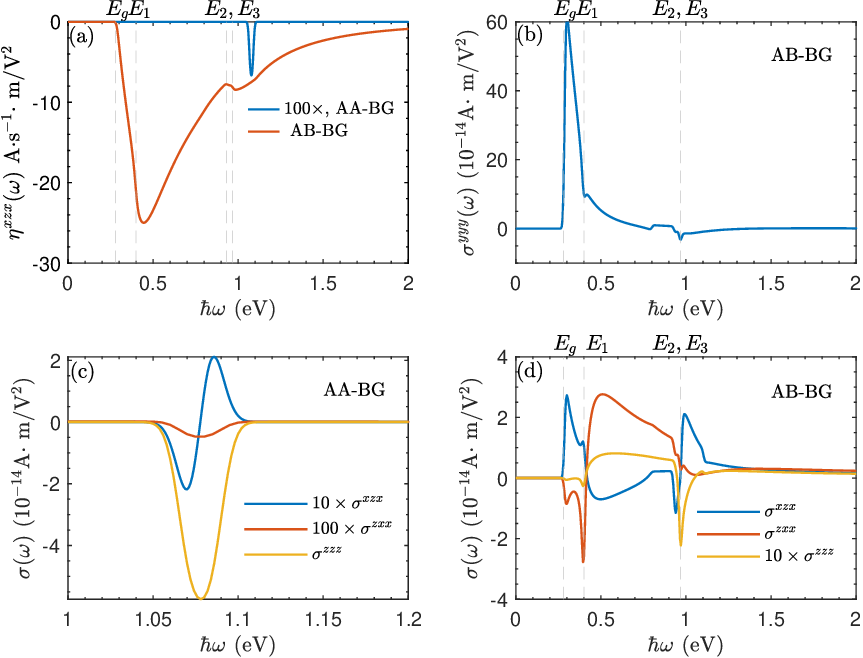}
	\caption{Injection coefficients and shift
          conductivities for AA-BG and AB-BG
          at $\Delta=0.4$~eV. (a) $\eta^{xzx}$ for AA-BG and
            AB-BG, (b) $\sigma^{yyy}$ for AB-BG, (c) $\sigma^{xzx}$, $\sigma^{zxx}$, and $\sigma^{zzz}$  for AA-BG, and (d)  $\sigma^{xzx}$, $\sigma^{zxx}$, and $\sigma^{zzz}$ for AB-BG. }
	\label{fig:0.4}
\end{figure*}

Figure~\ref{fig:0.4}\,(a) shows the  injection
  coefficient spectra for AA-BG and AB-BG. 
For the injection in AA-BG, the spectrum is just a peak located
  in a very narrow energy range $1.069$~eV$<\hbar\omega<1.087$~eV with
  an absolute value about $0.067$ $\rm A\cdot s^{-1}\cdot m/V^2$. From
  the analytic results shown in Eq.\,(\ref{eq:aa_ex-1}), the spectra include two
  contributions at different photon energy regions: one is from the optical transition between the bands (1, 3) for photon energy $\hbar\omega>2\sqrt{\Delta^2+\gamma_1^2}$ or $1.078$~eV$<\hbar\omega<$1.087~eV, and
  the other is between the bands (2, 4) for
  $\hbar\omega<2\sqrt{\Delta^2+\gamma_1^2}$ or $1.069$~eV$<\hbar\omega<$1.078~eV; both magnitudes are
  nearly proportional to
  $\hbar\omega-2\sqrt{\Delta^2+\gamma_1^2}$. These two contributions
  merge as a single peak just because the $\delta$ function is numerically
  broadened with $\Gamma=10$~meV, which is even
  larger than each energy region.
The injection coefficient $\eta^{xzx}$ in AB-BG starts with photon energy higher than the gap, i.e.,
  $\hbar\omega>0.28$~eV,  and reaches its maximum
value of $25~\rm A\cdot s^{-1}\cdot m/V^2$ in amplitude at
$\hbar\omega=0.45$~eV, which is slightly larger than the first
  VHS energy of JDOS  $E_1$; the
  energy difference arises from the zero electron velocity at this VHS. 
Considering the thickness of a bilayer graphene as $2c=6.7$ \AA,  the effective bulk injection coefficient is $3.7\times10^{10}~\rm \mu A\cdot s^{-1}V^{-2}$, which is nearly 50 times larger that that in bulk GaAs \cite{Rangel2017}.
  After this peak, the amplitude of injection coefficient
  decreases as the photon energy increases, except for a small peak located around the JDOS peak at higher energy $E_2$ or $E_3$. It can be seen that the
  injection coefficient for AB-BG is about two orders of magnitude
  larger than that for AA-BG.
To have a direct impression on these values, we give an estimation on how large the injection current can be in AB-BG.
Based on the Eq.\,(\ref{eq:inj_curren}), when the laser is
  a $45^\circ$ obliquely incident $p$-polarized light with photon
 energy of 0.45~eV, light intensity of $I=0.1~\rm GW/cm^2$, and pulse duration of $\tau=1$~ps, the generated injection current is
 $2\eta^{xzx}\frac{I}{2c\epsilon_0}W\tau\sim 9~\rm mA$ for an electrode with a width $W=1~\rm \mu m$.

Then we turn to the shift conductivities, as
  shown in Figs.\,\ref{fig:0.4}\,(b--d).
Figure~\ref{fig:0.4}\,(c) gives the shift conductivity for
AA-BG. It can be seen that the component $\sigma^{zzz}$ is
  about one order of magnitude larger than  
$\sigma^{xzx}$, or is at least two order of magnitude larger than $\sigma^{zxx}$. Both
  $\sigma^{zzz}$ and $\sigma^{xzx}$ have nonzero values only in the
  very narrow energy regions, similar to the 
  injection coefficient. These results are consistent with the
  analytic results shown in Eqs.\,(\ref{eq:aa_ex-2}--\ref{eq:aa_ex-3}).  Interestingly, 
$\sigma^{xzx}$ includes the contributions from the band 1 to 3 and
from the band 2 to 4 but with opposite signs.
For  AB-BG shown in Figs.\,\ref{fig:0.4}\,(b) and (d), all nonzero components start from the band
  edge $\hbar\omega\ge E_g$. Different from the injection
  coefficients, the shift conductivities at the band edge are
  nonzero, and show prominent peaks. Especially, $\sigma^{yyy}$
shows a large value about $6\times10^{-13}~\rm A\cdot m/V^2$ at the
  band edge and it drops quickly with increasing the photon energy.
The effective bulk shift conductivity is $896~\rm \mu A/V^2$, which is several times larger than in GeSe ($200~\rm \mu A/V^2$) \cite{Rangel2017}.
Besides, the component $\sigma^{zzz}$ is at least one order of
  magnitude smaller than other nonzero components, totally different
  from the case of AA-BG, where it is the largest one.
The spectra of $\sigma^{xzx}$ and $\sigma^{zxx}$ have similar
  amplitude around a few $10^{-14}~\rm A\cdot m/V^2$, which is a few
  tens of times smaller than the peak of $\sigma^{yyy}$; they also show
  some fine structures around those characteristic energies $E_1$, $E_2$,
  and $E_3$.
We repeat the above estimation for the shift current using the same parameters but $\hbar\omega=0.3$~eV, and then obtain the generated shift current of $2\sigma^{yyy}\frac{I}{2c\epsilon_0}W\sim 0.23$~mA.

\subsection{Effects of Gate voltage}

\begin{figure*}[ht]
	\centering
	\includegraphics[scale=1]{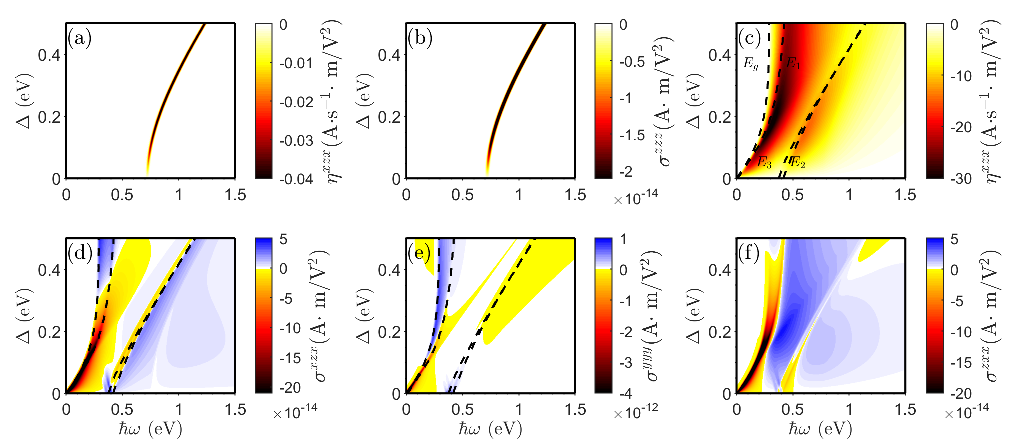}
	\caption{Gate voltage dependence of injection  coefficients and shift
          conductivities at zero chemical potential. (a)
          $\eta^{xzx}(\omega)$ and (b) $\sigma^{zzz}(\omega)$ for
          AA-BG, (c) $\eta^{xzx}(\omega)$, (d) $\sigma^{xzx}(\omega)$,
          (e) $\sigma^{yyy}(\omega)$, and (f) $\sigma^{zxx}(\omega)$
          for AB-BG. In (c--g), dashed curves indicate the characteristic
            energies $E_g$, $E_1$, $E_2$, and $E_3$ for AB-BG.}
	\label{fig:E}
\end{figure*}

Figure~\ref{fig:E} gives the gate voltage dependence of the
  injection coefficients and shift conductivities for AA-BG and AB-BG
  at zero chemical potential.
Note that the negative gate voltage leads to opposite
coefficients, which are consistent with the results
by Xiong \textit{et al.}~\cite{2DMaterials_8_35008_2021_Xiong}, thus only positive
  gate voltages are shown here.

Figures \ref{fig:E}\,(a) and (b) show the spectra of $\eta^{xzx}$
  and $\sigma^{zzz}$ for AA-BG, respectively. As indicated in previous section, both spectra for different gate
  voltages are nonzero in a very
  narrow photon energy region. With the increase of the gate voltage,
  the region moves to larger energy and the values of both spectra
  increase, which are indicated by $\propto\Delta$ in Eqs.\,(\ref{eq:anaresAA-1}) and (\ref{eq:anaresAA-3}).
Figure \ref{fig:E}\,(c) gives the injection coefficient
  $\eta^{xzx}$ for AB-BG. At each gate voltage, the injection
  coefficient shows two peaks located at photon energies slightly larger than 
  $E_1$ and $E_2$, which have been discussed in previous section. As the gate
  voltage $\Delta$ varies, the peak amplitude reaches 
  a maximum at $\Delta\sim0.2$~eV.
The shift conductivities $\sigma^{xzx},
  \sigma^{yyy}$ and $\sigma^{zxx}$ for AB-BG are plotted in
Figs.\,\ref{fig:E}\,(d--f). They show some similar
characteristics: (1) The spectra are located at about the band
  gap  similar to the case of $\Delta=0.4$~eV, and their amplitudes increase with the decrease of $\Delta$;  $\sigma^{xzx}$ and $\sigma^{zxx}$
  increase much faster than  $\sigma^{yyy}$. 
  (2) There exist sign changes of shift
  conductivities. 
\subsection{Effects of Chemical potential}

\begin{figure*}[ht]
	\centering
	\includegraphics[scale=1]{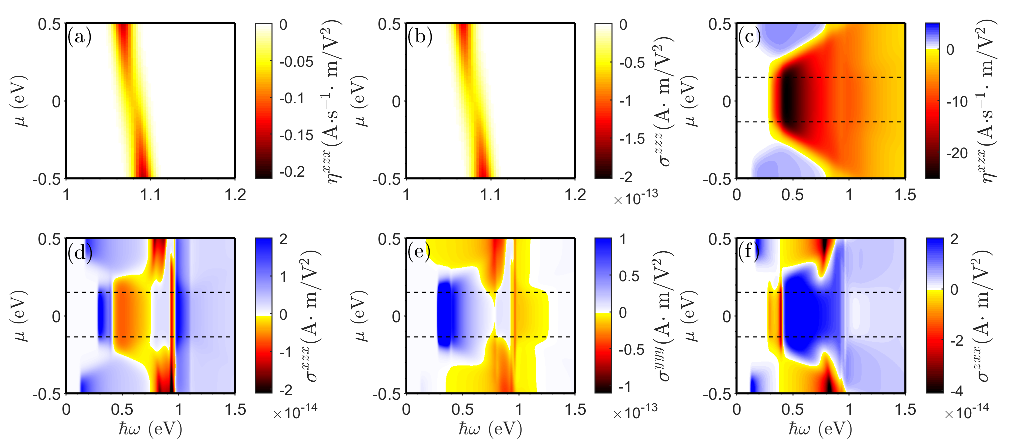}
	\caption{Chemical potential dependence of injection
          coefficients and and shift conductivities at
            $\Delta=0.4$~eV. (a)
          $\eta^{xzx}(\omega)$ and (b) $\sigma^{zzz}(\omega)$ for
          AA-BG, (c) $\eta^{xzx}(\omega)$, (d) $\sigma^{xzx}(\omega)$,
          (e) $\sigma^{yyy}(\omega)$, and (f) $\sigma^{zxx}(\omega)$
          for AB-BG. The dashed lines in (c--f) indicate the
            position of the conduction and valence band edges for AB-BG.}
	\label{fig:mu}
\end{figure*}

The chemical potential $\mu$ dependence of
   injection coefficients and shift conductivities
at $\Delta=0.4$~eV are depicted in Fig.\,\ref{fig:mu} with
  the same layout as Fig.\,\ref{fig:E}.
For AA-BG in Figs.\,\ref{fig:mu}\,(a) and (b), they
show very similar asymmetric dependence on the chemical
potential: with the increase of the chemical
potential, the values of all coefficients increase and the locations shift to higher or lower photon
energies depending on the sign of the chemical potential. For positive
chemical potential, the transitions between bands (1, 3) are
suppressed according to the Pauli blocking effects, while new extra transitions
between bands (2, 4) appear due to the additional free electrons in the
band 2. The extra transitions require lower photon energy and red
shift the spectra, and they also correspond to larger JDOS, leading to larger coefficients. Similar results can be
analyzed for negative chemical potential, but with switching the
band pairs (1, 2) and (3, 4).

In AB-BG, the chemical potential $\mu$ has different effects, as shown in Figs.\,\ref{fig:mu}\,(c--f). Due to the
  existence of the  band gap, the spectra are hardly changed when the
  chemical potential lies in the gap.
When $\mu$ is above the conduction band
  edge or below the valence
  band edge, the main
  peak of $\eta^{xzx}$ around 0.5 eV is reduced
gradually due to the Pauli blocking, and there appear new transitions between
  the bands (1, 2) or (3, 4) to give additional injections with opposite
  signs. Similar
  results are obtained for the shift conductivities.

\section{Conclusion}
In this paper we have studied the gate voltage induced injection current and shift current in AA- and AB-stacked bilayer graphene. 
The gate voltage plays a crucial role in breaking the inversion symmetry of bilayer graphene to induce photogalvanic effects, and at the same time it effectively changes the band structure for AB-BG with 
opening gaps located in the K-M directions and inducing additional VHS located in the K-$\Gamma$ directions. 
In AA-BG, the injection and shift currents are mainly induced by optical transitions between two pairs of nearly parallel bands; the coefficient spectra locate in a very narrow photon energy region of about 20~meV. 
In AB-BG, the optical transition can occur between any possible band pairs, and the structure of spectra are strongly determined by the band gap and the VHS energies.
For both structures, the injection and shift currents can be generated by the existence of an oblique $p$-polarized light component, while the in-plane shift currents in AB-BG can also be generated by normal incident lights. The out-of-plane shift current finally results in a static electric polarization between layers. The stacking order has significant effects on both currents. The  injection coefficient for AA-BG is about two orders of magnitude smaller than that for AB-BG, while the shift conductivities are mostly in the same order of magnitude.
All these coefficients can be effectively modulated by the gate voltage
and the chemical potential.
Our results suggest that gate voltage controlled bilayer graphene
  can be used to realize tunable optoelectronic detectors working in
  the mid-infrared.

\acknowledgments
This work has been supported by National Natural Science Foundation of China Grant No. 12034003, 12004379,  and 62250065. J.L.C. acknowledges the support from Talent Program of CIOMP.

\appendix
\section{Berry connections of AA-BG\label{app:xi}}
The general expression for the Berry connection of AA-BG is
\begin{align}
  \bm\xi_{nm\bm k}  &=\left(\sqrt{1-\alpha_n{\cal N}_{\beta_n\bm k}}
     \sqrt{1-\alpha_m{\cal N}_{\beta_m\bm k}} +\alpha_n\alpha_m\sqrt{1+\alpha_n{\cal N}_{\beta_n\bm k}}
     \sqrt{1+\alpha_m{\cal N}_{\beta_m\bm k}}\right)\times\notag\\
  &\times\frac{1+\beta_n\beta_m}{8}\left[\hat{g}_{\bm k}^\ast(i\bm\nabla_{\bm k}\hat{g}_{\bm k}) + \hat{\bm
		y} d\right] \notag\\
  &+ \left(\alpha_m\sqrt{1-\alpha_n{\cal N}_{\beta_n\bm k}}
	\sqrt{1+\alpha_m{\cal N}_{\beta_m\bm k}} +\alpha_n\sqrt{1+\alpha_n{\cal N}_{\beta_n\bm k}}
	\sqrt{1-\alpha_m{\cal N}_{\beta_m\bm k} }\right)\notag\\
  &\times\frac{\beta_n\beta_m-1}{8}\left[\hat{g}_{\bm k}^\ast(i\bm\nabla_{\bm k} \hat{g}_{\bm k}) - \hat{\bm
    y} d\right]\notag\\
  &+\frac{i\delta_{\beta_n\beta_m}}{2}\left( \sqrt{1-\alpha_n{\cal
    N}_{\beta_n\bm k}}\bm\nabla_{\bm k} \sqrt{1-\alpha_m{\cal
           N}_{\beta_m\bm k}}+  \alpha_n\alpha_m\sqrt{1+\alpha_n{\cal
           N}_{\beta_n\bm k}}\bm\nabla_{\bm k} \sqrt{1+\alpha_m{\cal
           N}_{\beta_m\bm k}}\right)\notag\\
  &+\left(\sqrt{1-\alpha_n{\cal
    N}_{\beta_n\bm k}} + \alpha_n\sqrt{1+\alpha_n{\cal
           N}_{\beta_n\bm k}} \right)\left(
    \sqrt{1-\alpha_m{\cal N}_{\beta_m\bm k}}+
    \alpha_m\sqrt{1+\alpha_m{\cal N}_{\beta_m\bm k}}  \right)\notag\\
  &\times \frac{1+\beta_n\beta_m}{8}c\hat{\bm z}
\end{align}
with $d=\sqrt{3}/3a_0$. Here we give the $x$-component between
different bands as
\begin{subequations}
\begin{align}
  r^{x}_{13\bm k}&=-r^{x}_{31\bm k}=-\frac{i}{2}\frac{\frac{\partial
                   {\cal N}_{-1\bm k}}{\partial k_x}}{\sqrt{1-{\cal
                   N}_{-1\bm
                   k}^2}}=-\frac{i}{2}\frac{\gamma_3}{|\Delta|}(1-{\cal
                   N}_{-1\bm k}^2)\frac{\partial |g_{\bm k}|}{\partial
                   k_x}\,,\label{eq:rx12}\\
  r^{x}_{24\bm k}&=-r^{x}_{42\bm k}=-\frac{i}{2}\frac{\frac{\partial
                   {\cal N}_{+1\bm k}}{\partial k_x}}{\sqrt{1-{\cal N}_{+1\bm k}^2}}=-\frac{i}{2}\frac{\gamma_3}{|\Delta|}(1-{\cal
                   N}_{+1\bm k}^2)\frac{\partial |g_{\bm k}|}{\partial
                   k_x}\,,\label{eq:rx34}\\
  r^x_{12\bm k}&=r^x_{21\bm k}=-r^x_{34\bm k}=-r^x_{43\bm k}\notag\\
                 &=\frac{1}{4}\left[\sqrt{1+{\cal N}_{-1\bm k}}
                   \sqrt{1-{\cal N}_{+1\bm k}} +\sqrt{1-{\cal N}_{-1\bm k}}
                   \sqrt{1+{\cal N}_{+1\bm k}}\right]\left[\hat{g}_{\bm
                   k}^\ast \left(i\frac{\partial \hat{g}_{\bm k}}{\partial k_x}\right)\right]\,,\\
  r^x_{32\bm k}&=-r^x_{23\bm k}=r^x_{14\bm k}=-r^x_{41\bm k}\notag\\
	&=\frac{1}{4}\left[\sqrt{1+{\cal N}_{-1\bm k}}
	\sqrt{1-{\cal N}_{+1\bm k}} -\sqrt{1-{\cal N}_{-1\bm k}}
	\sqrt{1+{\cal N}_{+1\bm k}}\right]\left[\hat{g}_{\bm k}^\ast
   \left(i\frac{\partial \hat{g}_{\bm k}}{\partial k_x}\right) \right]\,.	
\end{align}
\end{subequations}
Combining with other quantities in Eqs.~(\ref{eq:rz}) and
(\ref{eq:xinn}), the injection coefficients and
the shift conductivities can be evaluated. For the latter use, we also
need
\begin{align}
  {\cal V}_{21\bm k}&=\frac{2\gamma_3}{\hbar}{\cal N}_{-1\bm
                      k}\frac{\partial |g_{\bm k}|}{\partial k_x}\,,\\
  {\cal V}_{43\bm k}&=\frac{2\gamma_3}{\hbar}{\cal N}_{+1\bm
                      k}\frac{\partial |g_{\bm k}|}{\partial k_x}\,.
\end{align}

\section{Analytical expressions of $\eta^{xzx}$, $\sigma^{xzx}$, and
  $\sigma^{zzz}$ in AA-BG under the linear dispersion
  approximation}\label{app:xiaa}
Here we give the analytic results for $\eta^{xzx}$ in
Eq.\,(\ref{eq:AA_inj}), $\sigma^{xzx}$ in Eq.\,(\ref{eq:AA_sigmaxzx}) and
$\sigma^{zzz}$ in Eq.\,(\ref{eq:AA_sigmazzz}) under the linear
dispersion approximation around the Dirac points. The term of $\sigma^{zxx}$
is not discussed due to its very small magnitude, as shown in Fig.\,\ref{fig:0.4}\,(c).

The integrands of
$\eta^{xzx}$, $\sigma^{xzx}$, and  $\sigma^{zzz}$ are functions of
$|g_{\bm k}|$, $\frac{\partial |g_{\bm k}|}{\partial k_x}$, and
$\frac{\partial ^2 |g_{\bm k}|}{\partial k_x^2}$, where all terms
involving $|g_{\bm k}|$ can be simplified by using the properties of
the $\delta$ function. The function $\delta(\hbar\omega_{nm\bm
  k}-\hbar\omega)$ is nonzero only for $|g_{\bm k}|=G_{nm}$ with
\begin{align}
  \gamma_3 G_{31}&=\gamma_1-\sqrt{\left(\frac{\hbar\omega}{2}\right)^2-\Delta^2}\,,\text{
  for } \hbar\omega \ge2\sqrt{\Delta^2+\gamma_1^2}\,, \\
  \gamma_3 G_{42}&=\sqrt{\left(\frac{\hbar\omega}{2}\right)^2-\Delta^2}-\gamma_1\,,\text{
          for } \hbar\omega \le 2\sqrt{\Delta^2+\gamma_1^2}\,.
\end{align}
Further we get
\begin{align}
  \left.\left({\cal N}_{-1\bm k}\right)\right |_{|g_{\bm k}|=G_{31}} = - \left.\left({\cal N}_{+1\bm
  k}\right)\right|_{|g_{\bm k}|=G_{42}}= - \sqrt{1-\left(\frac{2\Delta}{\hbar\omega}\right)^2}\,.
\end{align}
\begin{enumerate}
\item By substituting the expressions of ${\cal V}_{nm\bm
    k}^x$, $r_{31\bm k}^x$, $r_{13\bm k}^z$, $r_{42\bm k}^x$, and
  $r_{24\bm k}^z$,  $\eta^{xzx}$ becomes
  \begin{align}
    \eta^{xzx}
    =& \frac{e^3}{2\pi \hbar^2}\int d\bm
      k \left(\frac{c\gamma_3^2}{2\hbar|\Delta|}\right)\left\{f_{12\bm k}{\cal
      N}_{-1\bm k}^2(1-{\cal N}_{-1\bm k}^2)\left(\frac{\partial |g_{\bm
      k}|}{\partial k_x}\right)^2\delta(\omega_{31\bm
      k}-\omega)\right.\notag\\
    &\left.+f_{34\bm k}{\cal
      N}_{+1\bm k}^2(1-{\cal N}_{+1\bm k}^2)\left(\frac{\partial |g_{\bm
      k}|}{\partial k_x}\right)^2\delta(\omega_{42\bm
      k}-\omega)\right\}\notag\\
    =&\frac{e^3c|\Delta|}{\pi \hbar^2
      (\hbar\omega)^2}\left[1-\left(\frac{2\Delta}{\hbar\omega}\right)^2\right]\left\{f_{13\bm k}|_{|g_{\bm k}|=G_{31}} {\cal
      F}_{31}(\omega) + f_{24\bm k}|_{|g_{\bm k}|=G_{42}} {\cal
      F}_{42}(\omega) \right\}\,,
  \end{align}
  with
\begin{align}
  {\cal F}_{nm}(\omega) &= \int d\bm k \left(\gamma_3\frac{\partial |g_{\bm k}|}{\partial
                   k_x}\right)^2 \delta(\hbar\omega_{nm\bm
                          k}-\hbar\omega)\,.
\end{align}
\item To get the result for $\sigma^{xzx}$, we use 
  \begin{align}
    \frac{\partial {\cal N}_{-1\bm k}}{\partial k_x} =
   (1-{\cal N}_{-1\bm
    k}^2)^{3/2} \frac{\gamma_3}{|\Delta|}\frac{\partial |g_{\bm k}|}{\partial k_x}
  \end{align}
  to get
  \begin{align}
    r_{31\bm k}^z\frac{\partial r_{13\bm k}^x}{\partial k_x} +r_{31\bm
    k}^x\frac{\partial r_{13\bm k}^z}{\partial k_x} =&
    \frac{ic}{4}(1+{\cal N}_{-1\bm k}^2)(1-{\cal N}_{-1\bm
    k}^2)^{3/2}\left(\frac{\gamma_3}{|\Delta|}\frac{\partial |g_{\bm
    k}|}{\partial k_x}\right)^2\notag\\
    &-\frac{ic}{4}{\cal N}_{-1\bm k}(1-{\cal
    N}_{-1\bm k}^2)\frac{\gamma_3}{|\Delta|}\frac{\partial^2 |g_{\bm
    k}|}{\partial k_x^2}\,.
  \end{align}
Similar expressions can be obtained for terms involving $\bm r_{32\bm
  k}$. Then we get
  \begin{align}
	\sigma^{xzx}
	=&\frac{ e^3c}{4\pi\hbar(\hbar\omega)^2}\left\{\left[2-\left(\frac{2\Delta}{\hbar\omega}\right)^2\right]\frac{2|\Delta|}{\hbar\omega}\left[f_{13\bm k}|_{|g_{\bm k}|=G_{31}}{\cal F}_{31}(\omega)+f_{24\bm k}|_{|g_{\bm k}|=G_{42}}{\cal F}_{42}(\omega)\right]\right.\notag\\
     &\left.-|\Delta|\sqrt{1-\left(\frac{2\Delta}{\hbar\omega}\right)^2}\left[f_{13\bm k}|_{|g_{\bm k}|=G_{31}}{\cal Q}_{31}(\omega)-f_{24\bm k}|_{|g_{\bm k}|=G_{42}}{\cal Q}_{42}(\omega)\right]\right\}\,.
\end{align} 
with
\begin{align}
		{\cal Q}_{nm}(\omega) &= \int d\bm k \gamma_3\frac{\partial^2 |g_{\bm k}|}{\partial k^2_x}\delta(\hbar\omega_{nm\bm k}-\hbar\omega)\,.
\end{align} 
\item The term of $\sigma^{zzz}(\omega)$  becomes
\begin{align}
	\sigma^{zzz}(\omega)=&\frac{e^3}{2\pi \hbar^2}\int d\bm k\left\{f_{13\bm k}\frac{c^2}{4}{\cal N}_{-1\bm k}^2c\sqrt{1-{\cal N}^2_{-1\bm k}}\delta(\omega_{31\bm
		k}-\omega)\right.\notag\\
	&\left.+f_{24\bm k}\frac{c^2}{4}{\cal N}_{+1\bm k}^2c\sqrt{1-{\cal N}^2_{+1\bm k}}\delta(\omega_{42\bm
		k}-\omega)\right\}\notag\\
	=&\frac{e^3c^3}{4\pi \hbar} \frac{|\Delta|}{\hbar\omega}\left[1-\left(\frac{2\Delta}{\hbar\omega}\right)^2\right]\left[f_{13\bm k}|_{|g_{\bm k}|=G_{31}}{\cal J}_{31}(\omega)+f_{24\bm k}|_{|g_{\bm k}|=G_{42}}{\cal J}_{42}(\omega)\right]\,.
\end{align}
with
\begin{align}
	{\cal J}_{nm}(\omega) &= \int d\bm k \delta(\hbar\omega_{nm\bm
		k}-\hbar\omega)\,.
\end{align} 
\end{enumerate}
When the optical transition occurs just around the Dirac points $\bm K$, we can
approximate $|g_{\bm k+\bm K}|=\sqrt{3}a_0k/2$, then the
$\delta$ functions can be worked out as 
\begin{align}
  \delta(2\sqrt{\Delta^2+(\gamma_3|g_{\bm k}|-\gamma_1)^2}-\hbar\omega)
  &=\frac{\delta\left(k-{2G_{31}}/(\sqrt{3}a_0)\right)}{\sqrt{3}a_0|\gamma_3|\sqrt{1-\left(\frac{2\Delta}{\hbar\omega}\right)^2}}\theta(\hbar\omega-2\sqrt{\Delta^2+\gamma_1^2})\,,\\
    \delta(2\sqrt{\Delta^2+(\gamma_3|g_{\bm k}|+\gamma_1)^2}-\hbar\omega)
  &=\frac{\delta\left(k-{2G_{42}}/(\sqrt{3}a_0)\right)}{\sqrt{3}a_0|\gamma_3|\sqrt{1-\left(\frac{2\Delta}{\hbar\omega}\right)^2}}\theta(2\sqrt{\Delta^2+\gamma_1^2}-\hbar\omega)\,.
\end{align}
Then we get 
\begin{align}
  \begin{pmatrix}{\cal J}_{31}(\omega) \\ {\cal J}_{42}(\omega)
  \end{pmatrix}
  &=\frac{8\pi
  }{3a_0^2\gamma_3^2{\sqrt{1-\left(\frac{2\Delta}{\hbar\omega}\right)^2}}}\left|\gamma_1-\sqrt{\left(\frac{\hbar\omega}{2}\right)^2-\Delta^2}\right|\begin{pmatrix}\theta(\hbar\omega-2\sqrt{\Delta^2+\gamma_1^2})\\\theta(2\sqrt{\Delta^2+\gamma_1^2}-\hbar\omega)
  \end{pmatrix}  \,,\\
   \begin{pmatrix}{\cal F}_{31}(\omega) \\ {\cal F}_{42}(\omega)
  \end{pmatrix} &=\frac{3a_0^2\gamma_3^2}{8} \begin{pmatrix}{\cal J}_{31}(\omega) \\ {\cal J}_{42}(\omega)
  \end{pmatrix}\,,\\
  \begin{pmatrix}{\cal Q}_{31}(\omega) \\ {\cal Q}_{42}(\omega)
  \end{pmatrix}
  &=-\frac{\pi
  }{{\sqrt{1-\left(\frac{2\Delta}{\hbar\omega}\right)^2}}}\begin{pmatrix}\theta(\hbar\omega-2\sqrt{\Delta^2+\gamma_1^2})\\\theta(2\sqrt{\Delta^2+\gamma_1^2}-\hbar\omega)
  \end{pmatrix}  \,,
\end{align}  
where two Dirac points have been counted in the integration.
In such
approximation, the expressions for $\eta^{xzx}$, $\sigma^{xzx}$, and $\sigma^{zzz}$ are expressed as 
  \begin{align}
	\eta^{xzx}(\omega)=&\frac{e^3c|\Delta|\sqrt{1-\left(\frac{2\Delta}{\hbar\omega}\right)^2}}{\hbar^2(\hbar\omega)^2
		}\left|\gamma_1-\sqrt{\left(\frac{\hbar\omega}{2}\right)^2-\Delta^2}\right|\left({\cal M}_{31}(\omega)+{\cal M}_{42}(\omega)\right)\,,\label{eq:aa_ex-1}\\
	\sigma^{xzx}(\omega)=
	&\frac{e^3c|\Delta|\left(\hbar^2\omega^2-2\Delta^2\right)}{2\hbar(\hbar\omega)^4\sqrt{1-\left(\frac{2\Delta}{\hbar\omega}\right)^2}}\left|\sqrt{1-\left(\frac{2\Delta}{\hbar\omega}\right)^2}-\frac{2\gamma_1}{\hbar\omega}\right|\left({\cal M}_{31}(\omega)+{\cal M}_{42}(\omega)\right)\notag \\
	&-\frac{ce^3|\Delta|}{4\hbar(\hbar\omega)^2}\left({\cal M}_{31}(\omega)-{\cal M}_{42}(\omega)\right)\,,\label{eq:aa_ex-2}\\
	\sigma^{zzz}(\omega)=&\frac{e^3c^3|\Delta|\sqrt{1-\left(\frac{2\Delta}{\hbar\omega}\right)^2}}{3\hbar(a_0\gamma_3)^2}\left|\sqrt{1-\left(\frac{2\Delta}{\hbar\omega}\right)^2}-\frac{2\gamma_1}{\hbar\omega}\right|\left({\cal M}_{31}(\omega)+{\cal M}_{42}(\omega)\right)\,,
	\label{eq:aa_ex-3}
\end{align}
respectively, with
\begin{align}
 \begin{pmatrix}{\cal M}_{31}(\omega) \\ {\cal M}_{42}(\omega)
\end{pmatrix}=\begin{pmatrix}f_{13\bm k}|_{|g_{\bm k}|=G_{31}}\theta(\hbar\omega-2\sqrt{\Delta^2+\gamma_1^2})\\
		f_{24\bm k}|_{|g_{\bm k}|=G_{42}}\theta(2\sqrt{\Delta^2+\gamma_1^2}-\hbar\omega)
	\end{pmatrix} \,.
\end{align}
Through the Taylor expansion, the above expressions around frequency $2\sqrt{\Delta^2+\gamma_1^2}$ can be approximated as
\begin{align}
	\eta^{xzx}(\omega)\approx&\frac{ce^3|\Delta|\left|2\sqrt{\gamma_1^2+\Delta^2}-\hbar\omega\right|}{8\hbar^2(\gamma_1^2+\Delta^2)}\left({\cal M}_{31}(\omega)+{\cal M}_{42}(\omega)\right)\,,\label{eq:anaresAA-1}\\
	\sigma^{xzx}(\omega)\approx&\frac{ce^3|\Delta|(2\gamma_1^2+\Delta^2)\left|2\sqrt{\gamma_1^2+\Delta^2}-\hbar\omega\right|}{32\hbar\gamma_1^2\sqrt{\gamma_1^2+\Delta^2}^3}\left({\cal M}_{31}(\omega)+{\cal M}_{42}(\omega)\right)\notag\\
	&-\frac{ce^3|\Delta|}{16\hbar(\gamma_1^2+\Delta^2)}\left({\cal M}_{31}(\omega)-{\cal M}_{42}(\omega)\right)\,,\label{eq:anaresAA-2}\\
	\sigma^{zzz}(\omega)\approx&\frac{ce^3|\Delta|\left|2\sqrt{\gamma_1^2+\Delta^2}-\hbar\omega\right|}{6\hbar a_0^2\gamma_3^2(\gamma_1^2+\Delta^2)}\left({\cal M}_{31}(\omega)+{\cal M}_{42}(\omega)\right)\,.\label{eq:anaresAA-3}
\end{align} 

\section{Eigenenergies of AB-BG \label{app:ab}}
The eigenenergies $\epsilon$ satisfy the equation
\begin{align}
  |H^{\text{AB}}_{\bm k} - \epsilon | = 0\,,
\end{align}
or
\begin{align}
  \epsilon^4 + x_2 \epsilon^2 + x_1 \epsilon + x_0 = 0\,,  \label{eq:epsab}
\end{align}
with
\begin{align}
  x_2 =& -{\gamma_1^\prime}^2-
        \left(2{\gamma_0^\prime}^2+{\gamma_3^\prime}^2+2{\gamma_4^\prime}^2\right)|g_{\bm k}|^2-2\left[\Delta^2+\left(\frac{\Delta^\prime}{2}\right)^2\right]\,,\\
  x_1=&-4{\gamma_0^\prime}{\gamma_4^\prime}\left({\gamma_1^\prime} |g_{\bm k}|^2+{\gamma_3^\prime}\text{Re}\left[g_{\bm
       k}^3\right]\right)+\Delta^\prime\left({\gamma_3^\prime}^2 |g_{\bm k}|^2 -
       {\gamma_1^\prime}^2\right)\,,\\
  x_0=&\left({\gamma_0^\prime}^2-{\gamma_4^\prime}^2\right)^2|g_{\bm
       k}|^4-2{\gamma_3^\prime}\left[{\gamma_1^\prime}\left({\gamma_0^\prime}^2+{\gamma_4^\prime}^2\right)-{\gamma_0^\prime}{\gamma_4^\prime}\Delta^\prime\right]\text{Re}[g_{\bm
    k}^3] \notag\\
  &+
    \left\{{\gamma_3^\prime}^2\left[{\gamma_1^\prime}^2+
       \Delta^2-\left(\frac{\Delta^\prime}{2}\right)^2\right]-\left(2{\gamma_0^\prime}^2-{\gamma_3^\prime}^2\right)\left[\Delta^2-\left(\frac{\Delta^\prime}{2}\right)^2\right]-2{\gamma_0^\prime}{\gamma_1^\prime}{\gamma_4^\prime}\Delta^\prime\right\}|g_{\bm
    k}|^2\notag\\  
  &+\left[\Delta^2-\left(\frac{\Delta^\prime}{2}\right)^2\right]\left[{\gamma_1^\prime}^2+
       \Delta^2-\left(\frac{\Delta^\prime}{2}\right)^2\right]\,.
\end{align}
Then the analytic expressions of the eigenenergies are 
\begin{align}
  \epsilon_{n\bm k} =
                      \frac{1}{2}\left[\alpha_n\sqrt{-2x_2-\beta_n\frac{2x_1}{\sqrt{y}}-y}+\beta_n\sqrt{y}\right]\,,\quad\text{
                      for } n=1, 2, 3, 4\,.
\end{align}
with
\begin{align}
  y &= \frac{1}{6}\left[4^{\frac{1}{3}}
  \left(y_1+\sqrt{y_1^2-4y_2^3}\right)^{\frac{1}{3}}+\frac{4^{\frac{2}{3}}y_2}{
  \left(y_1+\sqrt{y_1^2-4y_2^3}\right)^{\frac{1}{3}}}-4x_2\right]\,,\\
  y_1&=2x_2^3+27x_1^2-72x_2x_0\,,\\
  y_2&=x_2^2+12x_0\,.
\end{align}
At the Dirac
points with $g_{\bm k}=0$, the four eigenenergies are $\pm
\Delta-\frac{\Delta^\prime}{2}$,
$\pm\sqrt{\Delta^2+\gamma_1^2}+\frac{\Delta^\prime}{2}$.

In general the electron-hole symmetry for AB-BG
is broken due to the nonzero of $\gamma_4^\prime$ and
  $\Delta^\prime$. However, we find that $\gamma_4^\prime$ and
  $\Delta^\prime$ have negligble effects on the optical transition
  between the bands (2, 3). With setting $\gamma_4^\prime=0$ and
  $\Delta^\prime=0$, the eigenvalues become
  \begin{align}
    \epsilon_{n\bm k} &= \alpha_n
    \frac{1}{\sqrt{2}}\sqrt{z_1+\alpha_n\beta_n \sqrt{z_2}}\,, \label{eq:approxeps}
  \end{align}
  with
  \begin{align}
    z_1&={\gamma_1^\prime}^2+2\Delta^2 +
    \left(2{\gamma_0^\prime}^2+{\gamma_3^\prime}^2\right)|g_{\bm k}|^2\,,\\
    z_2&=4{\gamma_0^\prime}^2\left[{\gamma_3^\prime}^2|g_{\bm k}|^4+2{\gamma_1^\prime}{\gamma_3^\prime}\text{Re}[g_{\bm
    k}^3]+({\gamma_1^\prime}^2+4\Delta^2)|g_{\bm k}|^2\right] + \left({\gamma_3^\prime}^2|g_{\bm
         k}|^2-{\gamma_1^\prime}^2\right)^2\,.
  \end{align}
Obviously, the electronic states become electron-hole symmetric.
  Using
  Eq.~(\ref{eq:approxeps}), we can have analytic discussion on the
  band gap $E_g$ and the  VHS for ${\cal J}_{32}$. Around the Dirac
  point $\bm K$, the approximation $g_{\bm
    k+\bm K}=-r e^{i\theta} $ can be adopted for $ \bm k=
  \frac{2r}{\sqrt{3}a_0}(\cos\theta \hat{\bm x} + \sin\theta \hat{\bm
    y})$.
  For zero $\Delta$, the zero energy of $\epsilon_{3\bm k}$
  can be directly found from Eq.\,(\ref{eq:approxeps}) at $r=0$ or 
  $r=r_0=-\frac{\gamma_1^\prime
    \gamma_3^\prime}{{\gamma_0^\prime}^2}$ and
  $\theta=(2n+1)\pi/3$. Therefore, there exist in total four
  degenerate zero energy points in one Dirac cone at $\Delta=0$;
  one is at this Dirac point, and the other three locate along the K-M
  directions.
  Furthermore, for small $r$, $\epsilon_{3\bm k}$ can be
  approximated by
  \begin{align}
    \epsilon_{3\bm k}^2 =\Delta^2 + c_2
   r^2 + c_3 \cos(3\theta)
   r^3 + c_4 r^4 \,, \label{eq:E3approx}
  \end{align}
  with
  \begin{align}
    c_2&={\gamma_3^\prime}^2-\frac{4{\gamma_0^\prime}^2\Delta^2}{{\gamma_1^\prime}^2}\,,\\
    c_3 &=    -
          \frac{2{\gamma_0^\prime}^2{\gamma_3^\prime}}{\gamma_1^\prime
          }\,,\\
    c_4&=
         \frac{{\gamma_0^\prime}^2}{{\gamma_1^\prime}^2}\left[{\gamma_0^\prime}^2-2{\gamma_3^\prime}^2
         +
         \frac{4\Delta^2(2{\gamma_0^\prime}^2-{\gamma_3^\prime}^2)}{{\gamma_1^\prime}^2}
         + \frac{16{\gamma_0^\prime}^2\Delta^4}{{\gamma_1^\prime}^4}\right]\,.
  \end{align}
  From Eq.~(\ref{eq:E3approx}) the band structure around the Dirac points has following features:
    \begin{enumerate}
    \item For nonzero $\Delta$, the energy $\epsilon_{3\bm k}$ at the Dirac point $\bm K$ is an extreme, and it is a local minimum
    (maximum)    as $c_2>0$  ($c_2<0$), which corresponds to
    $|\Delta|<\Delta_c$ ($|\Delta| > \Delta_c$) with
    $\Delta_c=|\gamma_3^\prime\gamma_1^\prime/(2\gamma_0^\prime)|=0.0229$~eV.
  \item  We first look at the case $|\Delta|>\Delta_c$ ($c_2<0$). For a fixed
    $\theta$, $\epsilon_{3\bm k}$ around the Dirac point $\bm K$ has one more
    local minimum located at $r=r_e(\cos3\theta)$ with
    \begin{align}
      r_e(\cos3\theta)= \frac{-3c_3\cos3\theta+ \sqrt{9c_3^2\cos^23\theta-32c_2c_4}}{8c_4}\,.
    \end{align}
    When $r$ is fixed and $\theta$ varies, $\epsilon_{3\bm k}$ has local maxima as  $\cos3\theta=1$ and local minima as
    $\cos3\theta=-1$.   When both $r$ and $\theta$ are considered,
    there exists a minimum at $r=r_e(-1)$ and
    $\theta=(2n+1)\pi/3$ (along the K-$\Gamma$ directions for integer $n$), and a VHS point at $r=r_e(1)$ and
    $\theta=2n\pi/3$ (along the K-M directions).
  \item For the case $|\Delta|<\Delta_c$ ($c_2>0$), $\epsilon_{3\bm k}$ has no  VHS point around the Dirac points but
    the minimum along K-$\Gamma$ directions still exists.
  \item Similar analysis can be applied to study the JDOS ${\cal
      J}_{42}={\cal J}_{31}$. After ignoring $\gamma_4^\prime$ and
    $\Delta^\prime$, $\epsilon_{4\bm k}-\epsilon_{2\bm k}$ has a local
    minimum at the $\bm K$ point, and there is no VHS in  ${\cal
      J}_{42}$. Therefore, $\gamma_4^\prime$ and
    $\Delta^\prime$ play a key role in forming a VHS in ${\cal
      J}_{42}$. 
  \end{enumerate}

\bibliography{ref}
\end{document}